\documentstyle[version2,aps]{revtex} 
\begin{document}                                                        
\draft                                                                  
\twocolumn
\widetext
\begin{title}
Hole pockets in the $t - J$ model
\end{title}

\author{R. Eder and Y. Ohta}

\begin{instit}
Department of Applied Physics, Nagoya University, Nagoya 464-01, Japan
\end{instit}

\begin{abstract}
We present an exact diagonalization study of the electron
momentum distribution $n(\bbox{k})$ small clusters 
of $t-J$ model for different hole concentrations and $t/J$.
Structures in $n(\bbox{k})$ which were previously interpreted as
a `large' Fermi surface are identified as originating from the
well known many-body backflow. To obtain reliable 
information about the true Fermi surface, we focus on the 
regime $t$$<$$J$, where the backflow effect is weak and suppress 
the formation of a bound state by introducing a density repulsion 
between holes. We find clear signatures of a Fermi surface which
takes the form of small hole pockets.
Comparison of the scaling of $n(\bbox{k})$ and that of the 
quasiparticle weight with $t/J$ suggests that these
pockets persist also for $t$$>$$J$.

\end{abstract} 
                                                                               
\pacs{74.20.-Z, 75.10.Jm, 75.50.Ee}
\narrowtext                  
\topskip8cm
\section{Introduction}
The unusual properties of high-temperature superconductors
have led to great interest in the physics of correlated electrons
near a Mott-Hubbard metal-to-insulator transition. 
Thereby a particularly intriguing problem is the
volume of the Fermi surface (FS) for the
slightly less than half-filled band:
should one model the doped insulator by a dilute gas of
quasiparticles corresponding to the
doped holes (this would imply that the volume of the
FS is proportional to the hole concentration)
or do all electrons take part in the formation of the Fermi surface,
so that its volume is identical to that of free electrons?
It is the purpose of this paper 
to present evidence  that for finite clusters of
$t$$-$$J$ model the first picture is the correct one:
the FS as deduced from the momentum distribution takes
the form of small hole pockets. The $t$$-$$J$ model reads:
\[
  H =                                                    
 -t \sum_{< i,j >, \sigma}                                         
( \hat{c}_{i, \sigma}^\dagger \hat{c}_{j, \sigma}  +  H.c. )
 + J \sum_{< i,j >}[\;\bbox{S}_i \cdot
 \bbox{S}_j
 - \frac{n_i n_j}{4}\;].
\]
The $\bbox{S}_i$ are the electronic spin operators, 
$\hat{c}^\dagger_{i,\sigma} =-c^\dagger_{i,\sigma}(1-n_{i,-\sigma})$
and the sum over $<i,j>$ stands for a summation       
over all pairs of nearest neighbors.
Various authors\cite{BoncaI,StephanHorsch,Ding} 
have computed the momentum distribution
$n_\sigma(\bbox{k})$$=$$\langle \hat{c}_{\bbox{k},\sigma}^\dagger
\hat{c}_{\bbox{k},\sigma} \rangle$ for the two-hole ground
state of small clusters of this model (corresponding to a nominal 
hole concentration of $\sim$$10$\%) and found it roughly consistent 
with a free-electron picture: $n(\bbox{k})$ is maximum at
$\bbox{k}$$=$$(0,0)$, minimum at $\bbox{k}$$=$$(\pi, \pi)$. 
It has become customary\cite{StephanHorsch} to cite this as 
evidence that already at such fairly low hole concentrations
the $t$$-$$J$ model has a free electron-like (`large') FS.
It is straightforward to see, however, that
this shape of $n(\bbox{k})$ is simply the consequence of 
elementary sum-rules and has no significance
for the actual topology of the FS\cite{comment}.
We have therefore performed a systematic study of the
$n(\bbox{k})$ for various doping levels and $t/J$.
 
\section{Single hole case}
\topskip0cm
As compared to the uniform value of $1/2$ for the
half-filled case, the introduction of only a single hole
changes $n(\vec{k})$ in a rather complex way. 
Fig. \ref{fig1} shows $n(\bbox{k})$
for the single hole-ground states
with momentum $k_0=(\pi/2,\pi/2)$ in the $16$-site cluster and 
momentum $k_0=(2\pi/3,0)$ in the $18$-site cluster.
The $\bbox{k}$ dependence of $n(\bbox{k})$
is roughly consistent with  free electrons,
i.e. $n(\bbox{k})$ is large near $(0,0)$ and small near
$(\pi,\pi)$. This structure, which simply ensures 
negative kinetic energy\cite{comment}, is less pronounced
the smaller $t/J$. The second characteristic feature are `dips'
at $\bbox{k}_0$ for the minority spin (i.e. the `hole spin')
and at $\bbox{k}_0 + (\pi,\pi)$, for both spin directions.
These dips are more pronounced for smaller $t/J$.
The question arises which of these features should be 
associated with the FS, i.e. do we have a `large' FS 
already for a single hole or is there a `hole pocket' at 
$\bbox{k}_0$? We note that the magnitude of the discontinuity
in $n(\bbox{k})$ has to be equal to the weight of the quasiparticle 
peak in the single particle spectral function, $Z_h$.
Since $Z_h$ has a pronounced\cite{Dagottoetal} (and therefore 
characteristic)
dependence on $t/J$, a potential FS discontinuity
must have the same characteristic dependence on $t/J$.
Then, the `depth' of the dip at $\bbox{k}_0$ can be estimated
by comparing with a symmetry equivalent $\bbox{k}$-point
i.e. for $\vec{k}_0=(\pi/2,\pi/2)$ we consider 
$\Delta_{dip}=n_\downarrow(-\pi/2,\pi/2) -n_\downarrow(\pi/2,\pi/2)$,
for $\vec{k}_0=(2\pi/3,0)$ we study
$\Delta_{dip}=n_\downarrow(0,2\pi/3)-n_\downarrow(2\pi/3,0)$.
In Fig. \ref{fig2} these differences are compared to $Z_h$ 
(obtained from the single particle
spectral function for momentum transfer $\bbox{k}_0$ at half-filling)
for various values of $t/J$. Obviously, $\Delta_{dip}=Z_h$ over the 
entire range of $t/J$, so that
that the dip clearly originates from the Fermi 
level crossing of the quasiparticle band,
i.e. we have a `hole pocket' at $\bbox{k}_0$.
On the other hand, differences $\Delta n(\bbox{k})$ across
the `large' FS always show the opposite
behaviour under a variation of $t/J$ as $Z_h$,
indicating that these drops in $n(\bbox{k})$ are unrelated to any 
FS crossing. This suggests to associate this
structure in $n(\bbox{k})$ with the well-known `backflow' 
for interacting Fermi systems\cite{Nozieres}. Such a strong backflow
effect is by no means surprising if we consider the change in the
single particle Greens function upon removing one electron 
with momentum $\bbox{k}_0$ near the FS:
naively one might expect that the only effect be the 
shift of the quasiparticle peak at $\bbox{k}_0$ 
from the photoemission to the inverse photoemission spectrum.
If this were true, however, the integrated photoemission 
weight (which equals the total number of electrons)
had decreased only by $Z_h\ll1$. Hence the bulk of 
spectral weight shift from photoemission to inverse photoemission
must occur for momenta  $\bbox{k}\neq\bbox{k}_0$, 
i.e. the strong backflow. What is remarkable 
is the wide spread of the backflow in $\bbox{k}$-space. 
This implies that for each individual $\bbox{k}$ the change of 
$n(\bbox{k})$ due to removal of an electron at $\bbox{k}_F$ 
is $\sim 1/N$, with $N$ the system size. 
For a small finite hole concentration $\delta$ it seems reasonable 
that the backflow contributions from the individual holes are
additive, so that the total backflow contribution would scale
with $\delta$.\\ 
What remains to be explained are the `satellite dips'
at $\bbox{k}_0 + (\pi,\pi)$. The most natural explanation are
antiferromagnetic spin correlations. To see this, let us
consider the case $t \ll J$, where the state
$1/\sqrt{2} \hat{c}_{\bbox{k}_0,\downarrow} |\Phi_0 \rangle$ 
(with $|\Phi_0\rangle$ the half-filled ground state)
to good approximation is an eigenstate.
For this state 
\begin{eqnarray}
\langle n_{\bbox{k},\downarrow} \rangle
&=& \frac{1}{2} (1 - \delta_{\vec{k},\vec{k}_0} )
- \frac{2}{3} S(\bbox{k} - \bbox{k}_0),
\nonumber \\
\langle n_{\bbox{k},\uparrow} \rangle
&=& \frac{1}{2} - \frac{4}{3} S(\bbox{k} - \bbox{k}_0)
+\frac{1}{N},
\end{eqnarray}
with $S(\bbox{q})$ the static spin structure factor of 
$|\Phi_0\rangle$. Since the latter is peaked sharply at 
$\bbox{q}= (\pi,\pi)$ we have a natural explanation for the 
`satellite dips', and it seams reasonable to adopt this 
explanation also for larger values of $t/J$.\\
Summarizing the results obtained so far, we may say that
the introduction of a single hole changes $n(\bbox{k})$
in a rather complex way: there is a dip at the momentum of the
hole, which originates from the Fermi level crossing
of the quasiparticle band and thus represents the 
`Fermi surface'. The dip is
superimposed over a smooth free-electron-like variation, 
the familiar many-body backflow. As a consequence of the small 
quasiparticle weight, this backflow is very pronounced in the 
$t-J$ model. For later reference
we note that the backflow contribution to $n(\bbox{k})$  
to good approximation is a function of $|k_x| + |k_y|$ only
(this is also confirmed by investigating 
$n(\bbox{k})$ for other $\bbox{k}_0$).
Finally, the strong antiferromagnetic spin correlations
produce dips also at $\vec{k}_0 + \vec{Q}$.\\
This shape of $n(\bbox{k})$ can be easily understood
by recalling\cite{spinbags} that the elementary excitations
near the Fermi energy are spin bags, where the hole is dressed by
antiferromagnetic spin fluctuations; a simple calculation
in terms of the string picture\cite{EderBecker}
reproduces the numerical results quantitatively.
\section{Two hole case}
We proceed to the ground state with two holes.
Various authors\cite{BoncaI,StephanHorsch,Ding} have found that 
the free electron-like variation of $n(\bbox{k})$ observed already 
for a single hole becomes more pronounced for this doping level,
and based on the criterion $n(\bbox{k})>1/2$\cite{StephanHorsch} the
`Luttinger Fermi surfaces' in Fig. \ref{fig3} would be assigned.
However, by the same arguments as for a single hole, these Luttinger 
Fermi surface are ruled out: Fig. \ref{fig4} compares the $t/J$ 
dependence of differences $\Delta n(\bbox{k})$
across the respective Luttinger FS to that of the
quasiparticle weight in the spectral function for the
two-hole ground state. $Z_h$ decreases sharply, the
$\Delta n(\bbox{k})$ increase monotonically with
$t/J$. The drop in $n(\bbox{k})$ upon crossing the large FS thus 
is obviously unrelated to any true Fermi level crossing. 
Instead, comparison with Fig. \ref{fig2} shows that
the $t/J$ dependence of the $\Delta n(\bbox{k})$ is very similar
to the backflow contribution for a single hole.
More precisley, if we assume that the backflow for the two holes 
is simply additive, we expect for the `large FS' differences
in the two-hole ground state:
\begin{equation}
\Delta n(\bbox{k}) = \Delta n_\uparrow^{(1h)}(\bbox{k}) +
\Delta n_\downarrow^{(1h)}(\bbox{k}) ,
\label{add}
\end{equation}
where $\Delta n_\sigma^{(1h)}(\bbox{k})$ are the
corresponding differences in the single hole ground state.
Fig. \ref{fig5} compares the `large FS' $\Delta n(\bbox{k})$
with the estimates obtained from (\ref{add}) by using the 
$\Delta n_\sigma^{(1h)}(\bbox{k})$ 
shown in Fig. \ref{fig2}. Both the magnitude and the 
$t/J$ scaling are predicted very well by (\ref{add}), 
which clearly suggests to associate 
the `large FS' with the backflow contribution.\\
The question then is: what is the true FS and 
how can we make it visible in $n(\bbox{k})$?
Fig. \ref{fig1} shows that for a single hole the true 
Fermi surface (i.e. the hole pocket ) is most clearly visible 
for large $J/t$. This is simply related to the fact that 
the quasiparticle weight is large in this parameter region 
(see Fig. \ref{fig2}). Since the $t/J$-scaling of $Z_h$ is 
essentially the same at half-filling and in the two hole ground 
state, (see Figs. \ref{fig2} and \ref{fig4}) we thus may expect to 
see the clearest FS signatures for large $J/t$ 
also in the two-hole ground state.\\
For more than one hole, however, we face an additional problem:
the strong interaction between the holes, which
manifests itself e.g. in a sizeable 
negative binding energy\cite{BoncaII,HaPo}.
An interacting state of two `quasiparticles' reads
\begin{equation}
|\Psi_0\rangle = \sum_{\bbox{k}}
\Delta(\bbox{k}) a_{\bbox{k},\uparrow}^\dagger
a_{-\bbox{k},\downarrow}^\dagger |vac\rangle.
\label{bound}
\end{equation}
Thus, whereas for a single hole we could fix the location of 
the pocket simply by choosing the total momentum,
in the two-hole ground state the
holes will be distributed over different momenta
with probability $\sim |\Delta(\bbox{k})|^2$ 
and one may not hope to observe any FS signature
unless $\Delta(\bbox{k})$ is well
localized in $\bbox{k}$-space, i.e. $\Delta(\bbox{k})\sim 
\delta_{\bbox{k},\bbox{k}_0}$ with the
quasiparticle ground state $\bbox{k}_0$.
This in turn necessitates that the interaction energy be
smaller than differences in single particle energy
between neighboring $\bbox{k}$-points,
i.e. weak interaction and sufficiently strong dispersion.
With this in mind we add a density interaction term
$H_V$$=$$V\sum_{<i,j>} n_i n_j$, to the Hamiltonian;
adjusting the parameter $V$, one may hope to
reach a situation, where $H_V$ to a certain degree `cancels' 
the intrinsic attractive interaction of the holes.
In addition, we include a small
next-nearest neighbor hopping term in the Hamiltonian,
so as to lift the unfavourable (near) degeneracy of the
quasiparticle dispersion
along the surface of the magnetic Brillouin zone;
we fix the value of the respective hopping 
integral to be $t'=-0.1t$.
For the $16$-site cluster this term has the additional advantage
that it breaks the spurious additional symmetry due to the
mapping to a $2^4$ hypercube and selects a unique 
two-hole groundstate with momentum $(0,0)$.\\
Let us stress the following: due to the addition of these
terms we are strictly speaking no longer
considering the original $t$$-$$J$ model. 
It seems quite plausible, however, that if a FS
exists at all, its volume should be changed neither by
changing the kinetic energy ($t'$-term) nor by introducing
an additional interaction ($V$-term). \\
To demonstrate the adjustment of $V$,
Fig. \ref{fig6} shows the variation of the
hole density correlation function
$g(\bbox{R})$$=$$(1/2) \sum_i \langle (1-n_{\bbox{R}_i})
(1- n_{\bbox{R}_i + \bbox{R}})\rangle$ 
in the two-hole ground states of the $16$ and $20$ site cluster 
with $V$
(due to a subtle but understandable pathology in its geometry,
analogous results cannot be obtained for the $18$-site cluster,
see Appendix). Its essentially identical behaviour in both clusters
clearly signals a change of the net 
interaction between the holes from attraction to repulsion. 
For intermediate values of $V$, on the other hand,
$g(\bbox{R})$ is quite homogeneous
indicating that the single particle delocalization energy 
of the holes dominates over their
interaction. Given this plus the large $Z_h$
we should therefore be in an optimal position for
observing the FS.\\
Figs. \ref{fig7} and \ref{fig8} show the single particle 
spectral function $A(\bbox{k},\omega)$ for $J/t$$=$$2$
and the momentum distribution for $J/t$$=$$1$ and $J/t$$=$$2$ 
(and the respective `optimal' $V$). In the spectral function,
the chemical potential $E_F$ is located near the top but within
a group of pronounced peaks, well separated from another such group 
in the inverse photoemission spectrum. There are pronounced peaks 
both immediately above and below $E_F$ which comprise 
the bulk of spectral weight for the respective momenta.
Corresponding to the well defined `quasiparticle peaks' in
the spectral function, $n(\bbox{k})$ exhibits a sharp variation: 
hole pockets at $(\pi,0)$ and $(0,\pi)$. They are superimposed 
over the familiar backflow contribution, which again has the generic 
free electron like form so as to ensure negative kinetic energy.
Fig. \ref{fig8} also gives the values of the quasiparticle weight for 
the `Fermi momenta'. For $(2\pi/5,3\pi/5)$ in the $20$-site cluster
the `quasiparticle peak' in the phtoemission spectrum (PES)
actually consists of two peaks with approximately 
equal weight; we consider these as a single `broadened' peak, so that 
the two weights should be added. This is supported by the good agreement 
with $Z_h$ at $(\pi/2,\pi/2)$ in the $16$-site cluster (where no splitting
occurs) and the reasonable agreement with the $Z_h$ deduced from the 
inverse photoemission spectrum (IPES) at $(\pi,0)$ (where no splitting 
occurs either). The `depth' of the pockets approximately equals
$Z_h$ and both quantities consistently
decrease with decreasing $J/t$. \\ 
A somewhat astonishing feature of these results is
the location of the pockets at $(\pi,0)$ rather than at 
$(\pi/2,\pi/2)$. This can be traced back to the point group 
symmetry of the ground two-hole ground state:
when the symmetry of the half-filled
ground state is $A_1$ (or $s$), that of the two-hole ground state is
$B_1$ (or $d_{x^2-y^2}$) and vice versa (the former situation is realized
in the $16$ and $18$-site cluster, the latter in the $20$-site cluster).
Addition of two holes thus always is equivalent to adding an object with
$d_{x^2-y^2}$-symmetry, which implies that the pair wave function
$\Delta(\bbox{k})$ in (\ref{bound}) should have this symmetry as well.
This in turn implies $\Delta(\bbox{k})=0$
for $\bbox{k}$ along $(1,1)$, so that
occupation of $(\pi,0)$ is favoured.\\
Fig. \ref{fig9} shows the `FS discontinuities' in the $4\times4$
cluster under a variation of the repulsion
strength $V$. They show maxima when the density correlation 
function is most homogeneous, precisely as one would
expect for Fermions with a variable interaction strength.
In the spectral function the reduction of the discontinuities as
$V\rightarrow0$ manifests itself by the reduction in intensity of 
the big IPES peak at $(\pi,0)$ and the appearance of small low energy
IPES peaks at the momenta next to $(\pi,0)$: the pockets
are `washed out'. We note that the $\Delta n(\bbox{k})$
across the `large' FS remains unaffected,
another indication that it is unrelated to low-energy physics.\\
A possible explanation for the hole pocket FS
would be spin-density-wave-type 
broken symmetry: although the ground states
under consideration are spin singlets,
this might be realized if the 
fluctuations of the staggered magnetization $M_S$ were slow 
as compared to the hole motion, so that
the holes move under the influence of an `adiabatically 
varying' staggered field.
A possible criterion for this situation would be 
$\tau_{tr}$$\cdot$$\omega_{AF}$$\ll$$2\pi$,
where $\tau_{tr}$ is the time it takes for a hole to transverse
the cluster and $\omega_{AF}$ is the frequency of fluctuations of
$M_S$. We estimate (for $J/t=2$) the group 
velocity of the holes from the dispersion of 
the `quasiparticle peak' in the PES spectrum and,
using the energies indicated by arrows in Fig. \ref{fig7}
for the $20$-site cluster and the peaks at
$(\pi/2,0)$ and $(\pi/2,\pi/2)$ for the $16$-site cluster,
we find $\tau_{tr}$$\simeq$$2\pi/0.5t$($2\pi/0.2t$) for the
$20$ ($16$)-site cluster.
Typical frequencies for fluctuations of $M_S$
can be obtained from its correlation function, which, up to a
constant, equals the dynamical spin susceptibility
for momentum transfer $(\pi,\pi)$; a rigorous lower bound
on $\omega_{AF}$ thus can be obtained by subtracting the 
ground state energy from the energy of the lowest state with total 
momentum $(\pi,\pi)$ and the same point group symmetry as the ground 
state. This gives  $\omega_{AF}$$>$$0.9t$($1.2t$) for the
$20$ ($16$) cluster, i.e. $\tau_{tr}\cdot \omega_{AF}$$>$$2\pi$.
`Almost static' N\'eel order thus can be ruled out as origin of the 
small FS, even for this fairly large value of $J/t$.\\
As an additional check we have introduced exchange
terms $J'$ between $2^{nd}$ and $3^{rd}$ nearest neighbors to
reduce the spin correlations and again optimized the repulsion
to enable `free' hole motion. 
Ground state properties of this (highly artificial)
model are summarized in Fig. \ref{fig10}:
the momentum distribution, hole density correlation function
and spin correlation function
$S(|\bbox{R}|)= exp(i \bbox{Q}\cdot\bbox{R}) 
\langle \bbox{S}_i \cdot \bbox{S}_{i+\bbox{R}}\rangle$ 
(with $\bbox{Q}=(\pi,\pi)$). The density correlation function is 
homogeneous (no charge ordering), the spin correlations decay rapidly 
(no long range antiferromagnetic or spiral ordering) but still there 
are unambiguous hole pockets in $n(\bbox{k})$. The only possible
conclusion is that it is only the large $Z_h$ which 
makes the pockets visible in the large $J$ region, and not the 
onset of any kind of ordering.\\
While the hole pockets can be made clearly visible for
large $J$, the situation is more involved for
$t$$>$$J$. In this parameter region
the small overlap between `quasiparticle'
and `bare hole' (as manifested by the small $Z_h$)
makes the $V$-term (which couples only to the bare hole)
increasingly inefficient in enforcing a noninteracting state:
rather than separating from each other, the two holes
remain bound on second-nearest neighbors up to fairly large 
values of $V$ and the `crossover' from attraction to
repulsion, which gave an unambigous prescription for choosing
$V$, cannot be obtained any more. We thus abandon both the $V$ and 
$t'$ terms and adopt a more indirect way of reasoning.\\
In the single hole case, we found that the pocket was superimposed 
over the smooth backflow contribution. We assume that the situation
for $2$ holes is similar, only with the additional complication that 
the pockets are now `washed out' due to the interaction between holes
(see Fig. \ref{fig9}). Therefore we expect that $n(\bbox{k})$ can 
be written as 
\begin{equation}
n(\bbox{k}) = n_{back}(\bbox{k}) 
+ Z_h \cdot |\Delta(\bbox{k})|^2,
\end{equation}
with the pair wave function $\Delta(\bbox{k})$ introduced 
in (\ref{bound}). As discussed above, the point group symmetry of the 
two-hole ground state necessitates that $\Delta(\bbox{k})$ has 
$d_{x^2-y^2}$ symmetry, so that the pockets are located at $(\pi,0)$.
Then, since the symmetry of the ground state is unchanged
by adding either the $t'$ or $V$ term, we conclude that
this should also hold true in the absence of these terms.\\ 
Since $n_{back}(\bbox{k})$ to good
approximation is a function of $|k_x| + |k_y|$ only
(see Fig. \ref{fig1}), this contribution can be eliminated by 
forming the difference of two momenta with (almost) equal 
$|k_x| + |k_y|$. Next, if we choose one of these momenta 
along (or near) the $(1,1)$ direction, where
the $d_{x^2-y^2}$-symmetry requires that 
$\Delta(\bbox{k})$ vanishes (or is small), and the other at
(or near) $(\pi,0)$ we should obtain
\begin{equation}
\Delta n(\bbox{k}) = Z_h \cdot |\Delta(\pi,0)|^2,
\end{equation}
so that, in contrast to the `large FS' differences indicated in 
Fig. \ref{fig3} this difference should scale with $Z_h$. To check this 
prediction, the $t/J$ dependence of various such differences is shown 
in Fig. \ref{fig11} and obviously, they are to excellent
approximation proportional to $Z_h$ over a wide range of $t/J$.
The scaling of $n(\bbox{k})$ with $t/J$ thus is completely
consistent with the assumptions\\ 
a) that there are washed out hole pockets at $(\pi,0)$,\\
b) that these are superimposed over the smooth backflow 
contribution, which is the sum of the backflows for the
two individual holes (see Fig. \ref{fig5}).

\section{Comparison with other numerical studies and experiment}

While the hole pockets are very clearly visible for large $J$,
the evidence for their existence in the
physical regime is of a more indirect character.
A comparison with other numerical calculations 
and experiments on high-temperature superconductors is 
therefore necessary.
As far as numerical studies on small clusters are concerned,
hole pockets and/or rigid band behaviour
upon doping are consistently suggested by most of the
available numerical calculations. For the $t-J$ model,
Poilblanc and Dagotto\cite{PoilblancDagotto}
studied the $A(\bbox{k},\omega)$ for single hole states 
and concluded that the two-hole ground state in the
$4\times 4$ cluster shows hole pockets at $(\pi,0)$, in agreement 
with the present result. On the other hand, 
Stephan and Horsch\cite{StephanHorsch} studied $n(\bbox{k})$
and $A(\bbox{k},\omega)$ for the two-hole ground state
and concluded that there is neither rigid 
band behaviour nor hole pockets. However, these authors
based their conclusions solely on the qualitative inspection of 
a rather limited data set, which is largely
irrelevant\cite{comment} for deciding the FS topology.
In addition to the inconsistent scaling behaviour
found above (Fig. \ref{fig4}), numerical calculation
of $A(\bbox{k},\omega)$ for the $20$-site cluster\cite{rigid} rules 
out the Luttinger FS postulated by Stephan and Horsch.\\ 
Castillo and Balseiro\cite{CastilloBalseiro} 
computed the Hall constant and found its
sign near half-filling to be consistent with a hole-like
FS, i.e. with hole pockets.
Gooding {\it et al.}\cite{Goodingetal} studied
the doping dependence of the spin correlation function
in clusters with special geometry and
also found indications of rigid-band behaviour. Finally, a
systematic study of the doping dependence of the single particle
spectral function\cite{rigid} shows rigid-band behaviour,
i.e. holes are filled into the quasiparticle band
present at half-filling (which naturally implies hole pockets).\\
The situation is quite similar for the Hubbard model.
While the generic\cite{comment} free-electron like shape 
of $n(\bbox{k})$
found in earlier Monte-Carlo studies\cite{Moreo1} was initially
considered as evidence against hole pockets,
more careful and systematic analysis\cite{Moreo2} showed that
hole pockets are in fact remarkably consistent with 
the numerical data, their nonobservation in the earlier 
studies being simply the consequence of thermal smearing.
It seems fair to say that the available numerical 
results for small clusters of both Hubbard and $t$$-$$J$ models,
when interpreted with care, are all consistent with
rigid band behaviour and/or hole pockets.\\
Let us next discuss experimental results on high-temperature
superconductors assuming that the hole pockets found in the
cluster studies persist in the real systems. 
The volume of the
FS associated with the $Cu-O$ plane-derived bands
in these materials presents a well-known puzzle: 
early photoemission experiments\cite{Olson}
show bands, which disperse towards the Fermi energy and vanish
at points in $\vec{k}$-space which are roughly
located on the free electron FS corresponding to
electron density $1-\delta$, where $\delta$ is the hole
concentration; on the other hand
transport properties can be modelled well\cite{Trugman,dagobert}
by assuming a FS with a volume $\sim \delta$.
In a Fermi liquid, the apparently contradicting 
quantities actually fall into distinct classes: 
photoemission spectra depend on $Z_h$,
transport properties do not. Hence, if one wants to resolve the
discrepancy entirely within a Fermi liquid-like picture,
the simplest way would be to assume a `small' FS
and explain the photoemission results by a
systematic variation of $Z_h$
along the band which forms the FS, similar to
the `shadow band' picture\cite{KampfSchrieffer}.
A trivial argument for such a strong $\bbox{k}$-dependence of the
quasiparticle weight is, that a distribution of
PES weight in the Brillouin zone 
(and hence a $n(\bbox{k})$) that resembles the
nointeracting FS, always optimizes the
expectation value of the kinetic energy. Therefore it is
favourable if those parts of the band structure,
which lie inside the free-electron FS have large spectral
weight, and the parts outside small weight.
Then, it seems that in a recent photoemission
study by Aebi {\em et al.}\cite{Aebi}, structures which are 
very consistent with such a shadow band scenario have indeed been 
observed. Moreover, another key feature of the dispersion relation
for a single hole, namely the extended flat region near 
$(\pi,0)$\cite{Maekawa,I}
has also been found as an universal feature of high temperature
superconductors\cite{Dessau,Abrikosov}
Adopting a rigid band/hole pocket scenario thus would explain
many experiments in a very simple and natural way,
which moreover is remarkably consistent with the
existing numerical data as a whole.

\section{Conclusion}

We have discussed the problems in directly determining the FS
from $n(\bbox{k})$ in the $t$$-$$J$ model:
small quasiparticle weight, a pronounced `backflow' effect 
and strong interaction between the doped holes. We have then 
examined the single particle spectral function
and $n(\bbox{k})$ in a situation where these problems
were largely avoided and found signatures of a FS 
which takes the form of small hole-pockets.
Analysis of the scaling of $n(\bbox{k})$ with $t/J$ 
suggested that these pockets also persist in the regime $t>J$. 
Available numerical data all support this picture, and we have
outlined a possible scenario to reconcile experiments on
high-temperature superconductors.\\
The assumption of a small Fermi surface implies that the phase of 
some given basis state is determined by a Slater determinant of rank 
$N-N_e$, ($N_e$ being the number of electrons) rather than $N_e$ 
(as it would be e.g. in a Gutzwiller projected Fermi sea). Moreover, 
it would not be the positions of the electrons which enter this 
Slater determinant, but those of the hole-like quasiparticles, 
so that we have a very different nature of long range phase 
coherence. The Fermi surface in an interacting system, being a 
`remnant' of the noninteracting one, is obviously a consequence
of the requirements to have minimum kinetic energy and to satisfy the
Pauli principle. On the other hand, close to half-filling most of 
the electrons are immobile, so that the gain in kinetic energy
from creating the long range phase coherence between
electrons (which is responsible for
the singularity of $n(\bbox{k})$) may not be very large.
On the other hand, the vacancies are almost unconditionally
mobile, so that phase coherence between holes may be more
favourable.

It is a pleasure for us
to acknowledge numerous instructive discussions with Professor 
S. Maekawa. Financial support of R. E. by the Japan Society for the 
Promotion of Science is most gratefully acknowledged.
Computations were partly carried out at the computer Center of the
Institute for Molecular Science, Okazaki National Research Institues.
\section{Appendix}
The $18$-site cluster has a pathological
geometry, which does not allow for an unbound state of two
particles with $d_{x^2-y^2}$-symmetry. The ultimate reason
is that the primitive lattice translations in this cluster
are $(3,3)$ and $(3,-3)$. Writing an interacting two-hole state 
in real space, we have
\begin{equation}
|\Phi\rangle = \sum_i \sum_{\bbox{R}} \phi(\bbox{R}) 
a_{i,\uparrow}^\dagger a_{i+\bbox{R},\downarrow}^\dagger
|vac\rangle
\end{equation}
If we choose $\bbox{R}=(2,1)$, rotate counterclockwise 
by $\pi/2$, reflect by the $x$-axis and add $(3,3)$ we recover the
original vector. A state with $d_{x^2-y^2}$ symmetry picks up a factor
of $(-1)$ during these operations, hence $\phi(2,1)=0$.
Analogous reasoning shows that $\phi(3,0)=0$, so that
all large distances between particles are `symmetry forbidden'
(possible distance in this cluster are $(1,0)$, $(1,1)$, $(2,0)$
$(2,1)$ and $(3,0)$).
The `unbinding transition' for this cluster thus can occur only
via level crossing and this is indeed the case: when $V$ is 
switched on in the $d_{x^2-y^2}$ ground state,
the holes stay close to each other even for fairly large values
of $V$. Instead, at $V\sim 3t$ (for $J/t=2$) a level crossing occurs, 
and a new ground state with momentum $(2\pi/3,0)$ is stabilized.

\figure{Momentum distribution for the single hole ground state 
        with $S_z=1/2$ (i.e. with a `$\downarrow$-hole')
        of the $4\times4$ cluster with $t/J=4$ (top)
        of the $4\times4$ cluster with $t/J=1$ (middle)
        $18$-site cluster with $t/J=2$ (bottom).
        The upper values refer to the majority spin, the lower
        values to the minority spin,
        the ground state momentum $\bbox{k}_0$ is marked by a black box
        and $\bbox{k}_0+(\pi,\pi)$ by a dotted box.
\label{fig1}}
\figure{Comparison of the $t/J$-dependence of $Z_h$ 
        at half-filling (dark squares)
        and various differences $\Delta n(\bbox{k})$ in the single hole
        ground states with $S_z=1/2$.
        Shown is the `depth' of the pockets (light circles)
        and differences across the `large FS' (up and down triangles). 
\label{fig2}}
\figure{Allowed momenta and `Luttinger Fermi surfaces' for
        various clusters.
        The `Fermi momenta' are denoted by $\bbox{k}_F$.
\label{fig3}}
\figure{Comparison of the $t/J$-dependence of $Z_h$
        (dark squares, obtained from the photoemission spectra at the 
        $\bbox{k}_F$ indicated in Fig. \ref{fig3})
        and the differences $\Delta n(\bbox{k})$  between pairs of
        momenta connected by dashed lines in Fig. \ref{fig3} 
        (light squares and circles).
\label{fig4}}
\figure{Comparison of the $t/J$-dependence of 
        the $\Delta n(\bbox{k})$ shown in Fig. \ref{fig3} 
        (light squares)  
        and the estimates obtained from (\ref{add})
        by using the single hole $\Delta n(\bbox{k})$
        in Fig. \ref{fig2} (dark squares).
\label{fig5}}
\figure{$V$-dependence of the hole density correlation 
        function $g(\bbox{R})$ 
        in the ground state of $t$$-$$t'$$-$$J$$-$$V$ clusters 
        with 2 holes ($J/t$$=$$2$).
\label{fig6} }
\figure{Single particle spectral function for
        the $t$$-$$t'$$-$$J$$-$$V$ model with 2 holes and $J/t$$=$$2$
        for the $20$-site with $V/t$$=$$2.5$ (a)
        as well as the $16$-site cluster with $V/t$$=$$2.4$ (b). 
        Delta functions have been replaced by
        Lorentzians of width $0.05t$.
        The frequency region $\omega$$<$$E_F$ ($\omega$$>$$E_F$) 
        corresponds to photoemission (inverse photoemission).
\label{fig7}}
\figure{Momentum distribution in the two-hole ground state
        of the cluster
        $t$$-$$t'$$-$$J$$-$$V$ models with 
        $J/t$$=$$2$ and 
        $V/t$$=$$2.5$ ($V/t$$=$$2.4$) for the $20$-site
        ($16$-site) cluster (a)
        and for $J/t$$=$$1$ and 
        $V/t$$=$$3.0$ ($V/t$$=$$2.0$) for the $20$-site
        ($16$-site) cluster (b).
        For the `Fermi momenta' 
        the quasiparticle weight $Z_h$ is given in brackets.
\label{fig8}}
\figure{$V$ dependence of selected $\Delta n(\bbox{k})$ in the
        $4\times 4$ cluster, parameters are like in Fig. \ref{fig6}.
\label{fig9}}
\figure{Hole density correlation function $g(\bbox{R})$ (dark circles), 
        spin correlation function $S(\bbox{R})$ (light squares)
        and $n(\bbox{k})$ (inset) for the ground state
        with spin frustration. Parameter values are $J=2$, $J'=0.75$,
        $t'=-0.1$. There is a density repulsion of strength
        $1.7t$ between holes on $1^{st}$, $2^{nd}$ and $3^{rd}$
        nearest neighbors.
\label{fig10}}
\figure{Comparison of the scaling of $Z_h$ (dark squares) and selected 
       difference $\Delta n(\bbox{k})$ (light squares) with $t/J$:
        $1.5$$\cdot$$(n(\pi/2,\pi/2)- n(\pi,0))$ ($16$-site),
        $2.8$$\cdot$$(n(\pi/3,\pi/3)- n(2\pi/3,0))$ ($18$-site) and
        $3.6$$\cdot$$(n(\pi/5,3\pi/5)- n(\pi,0))$ ($20$-site).
\label{fig11}}
                                                        
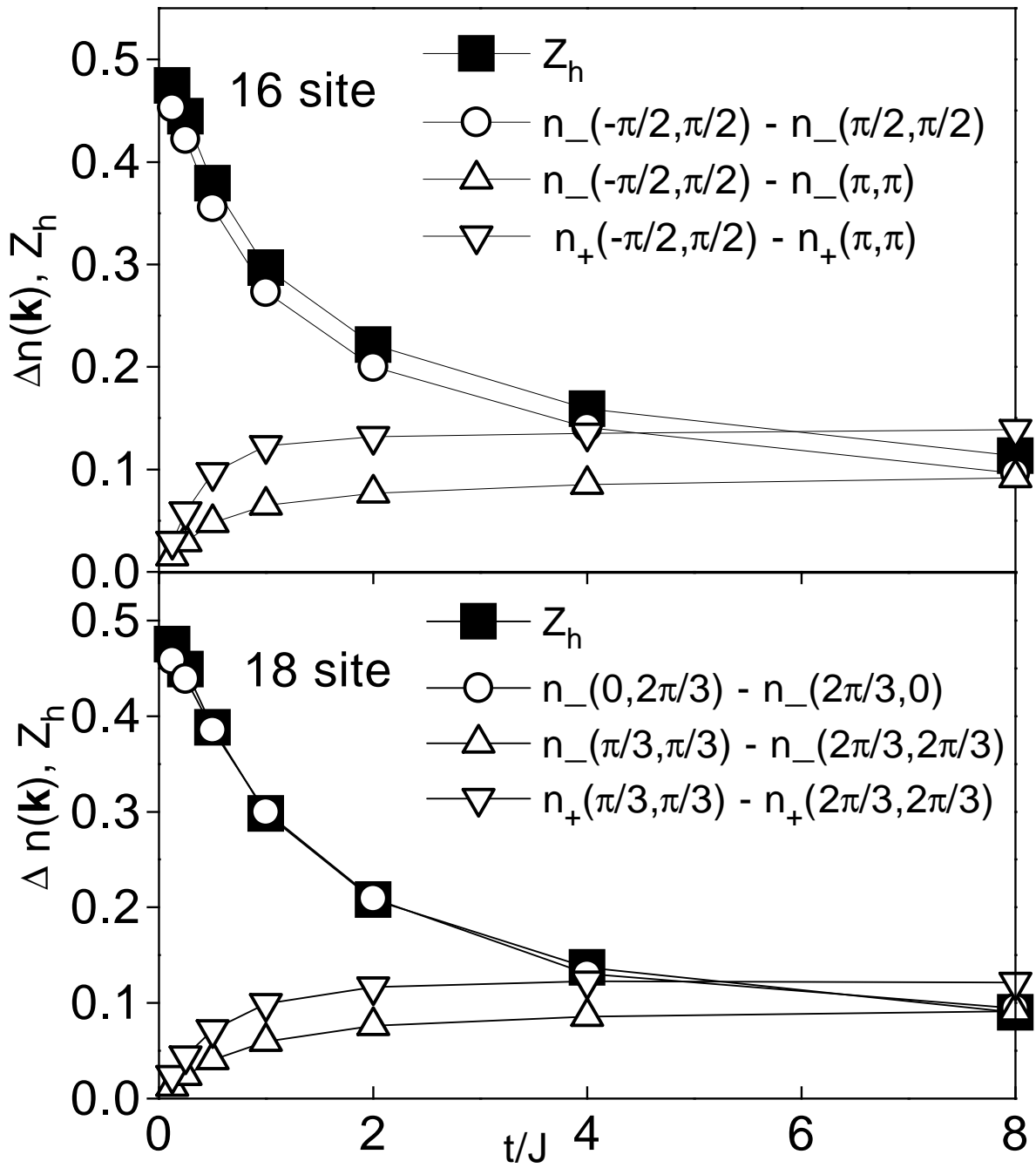
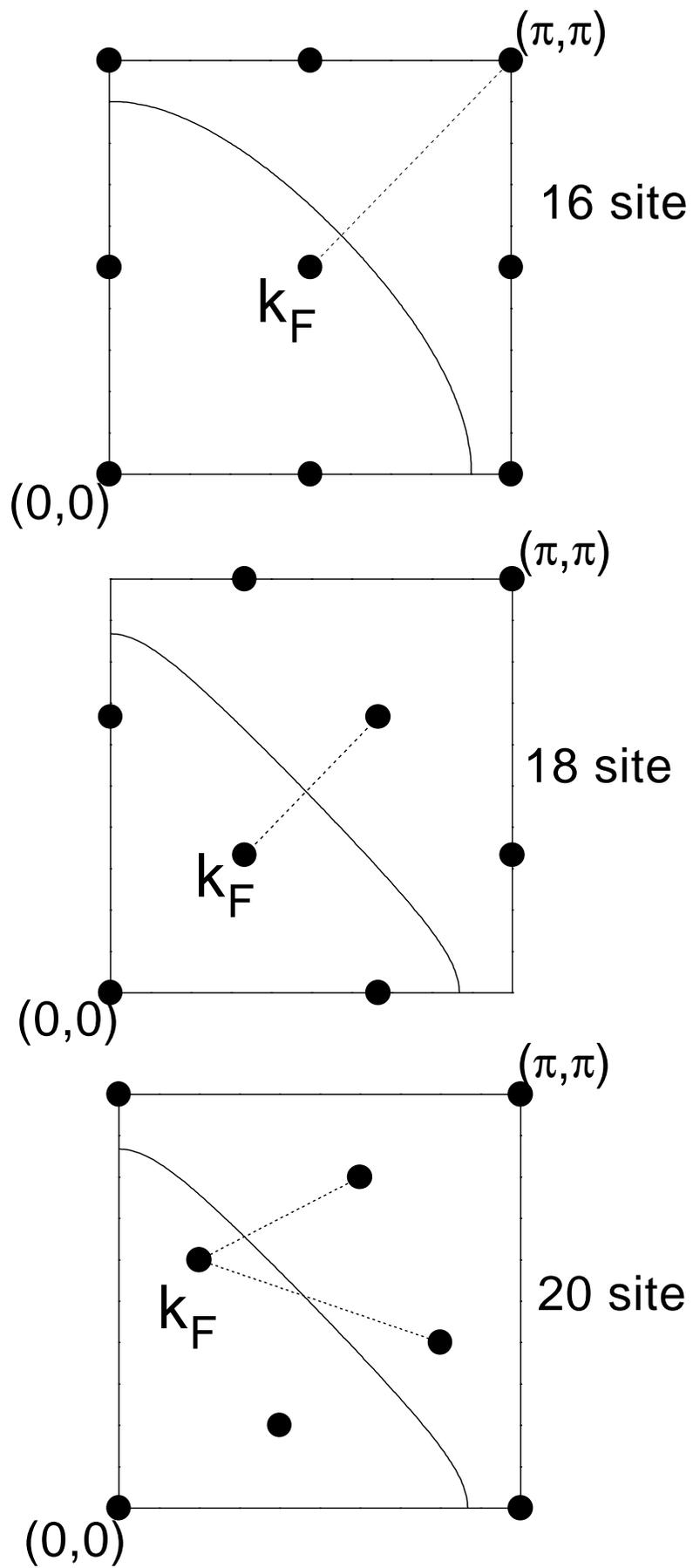
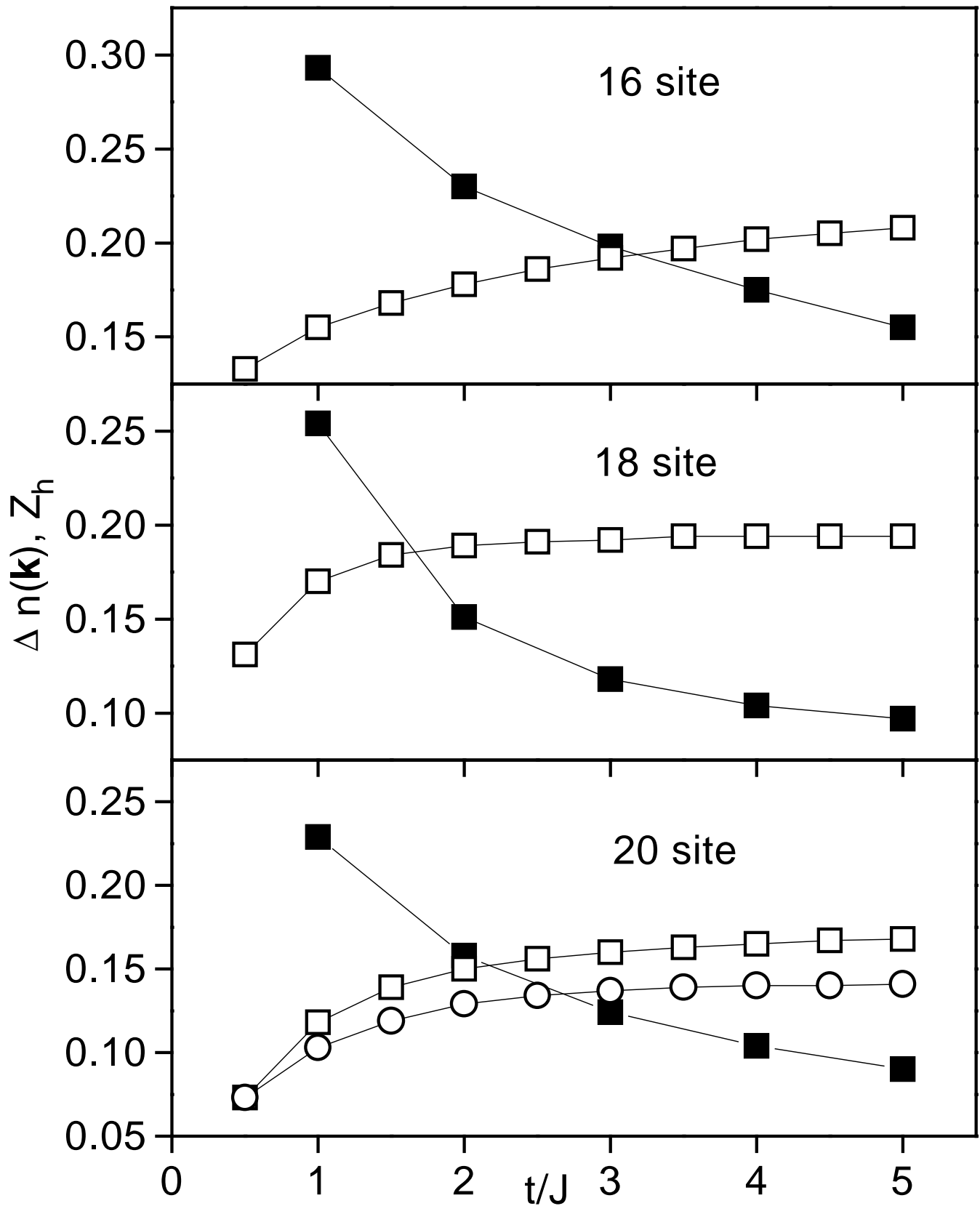
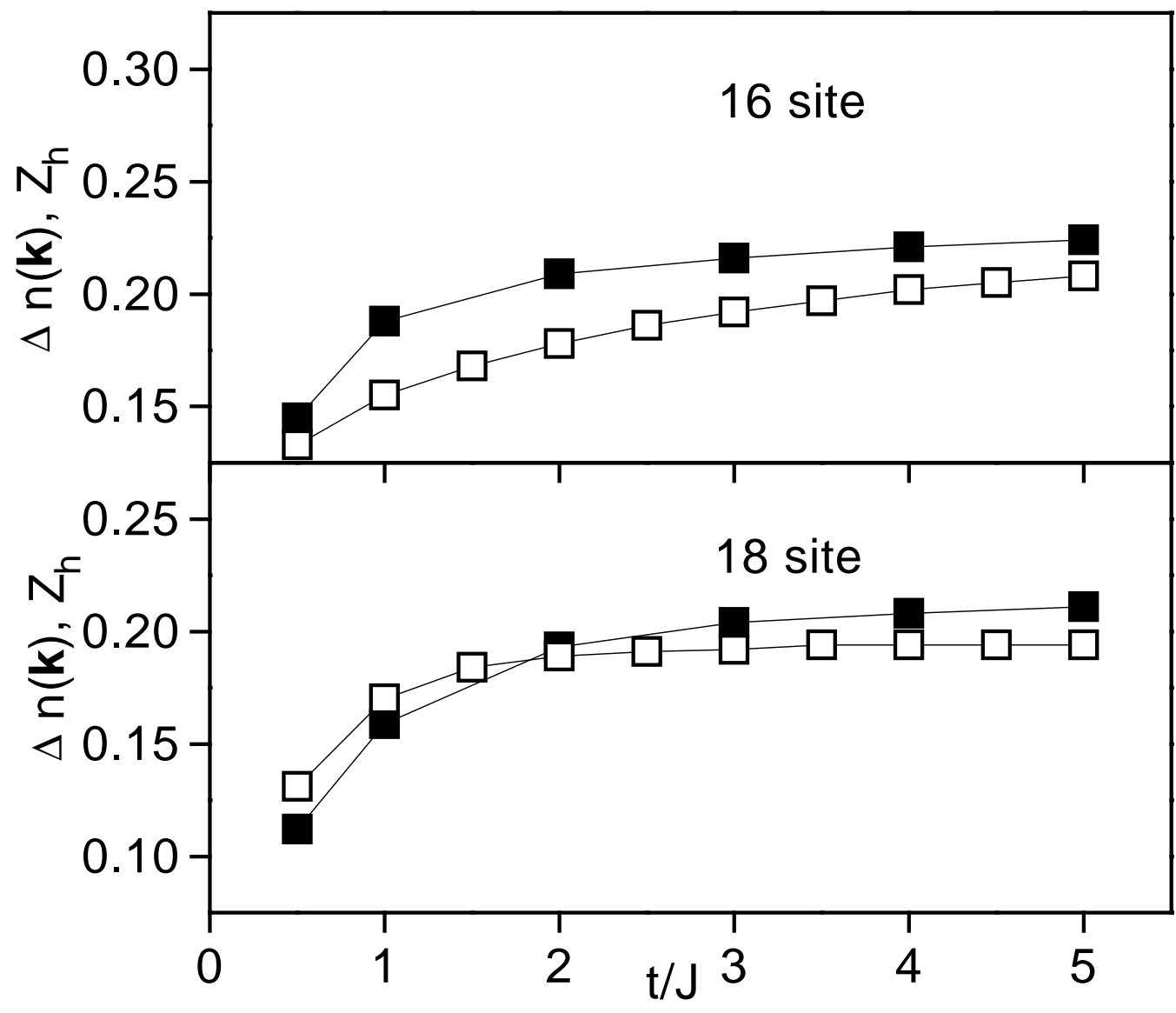
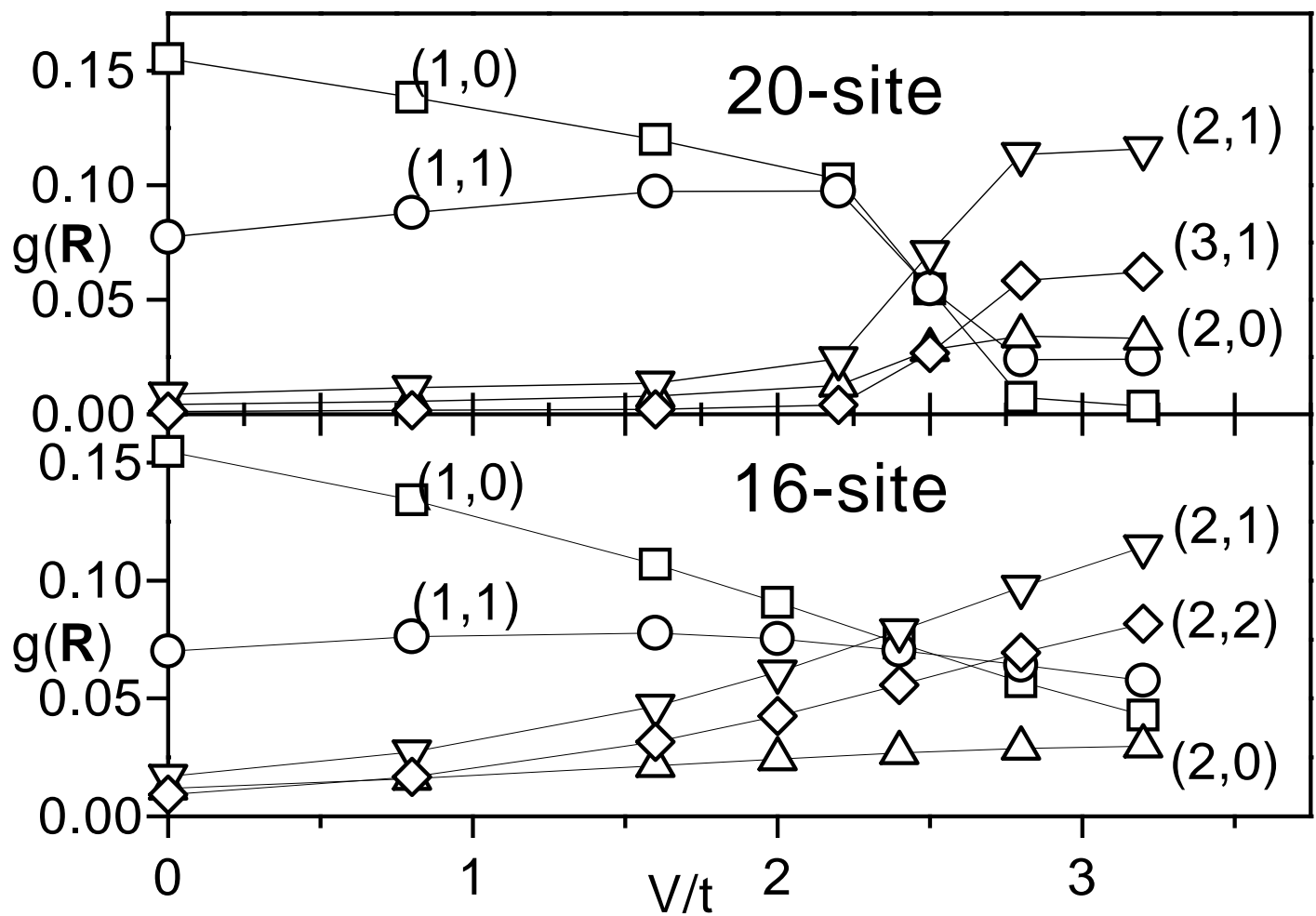
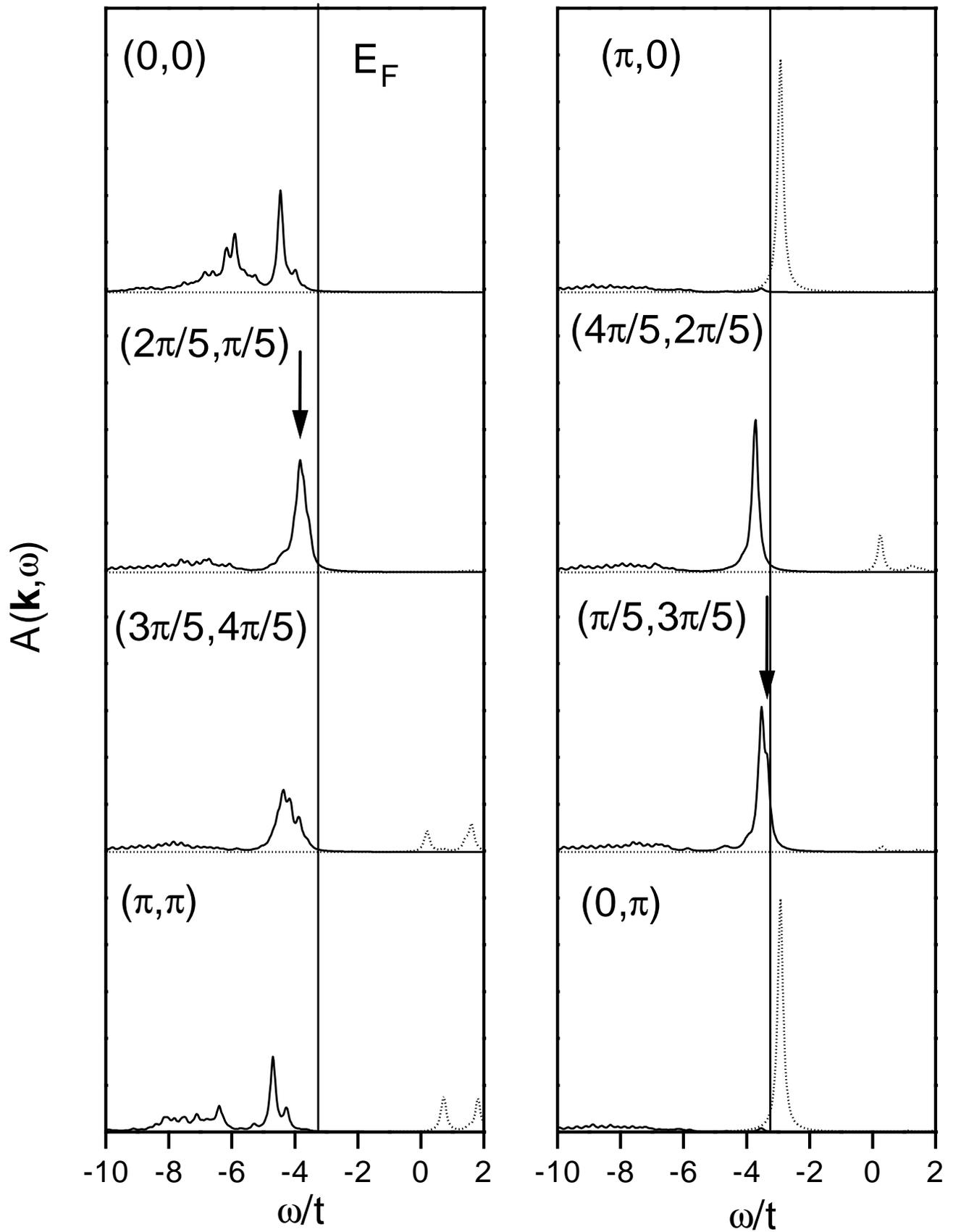
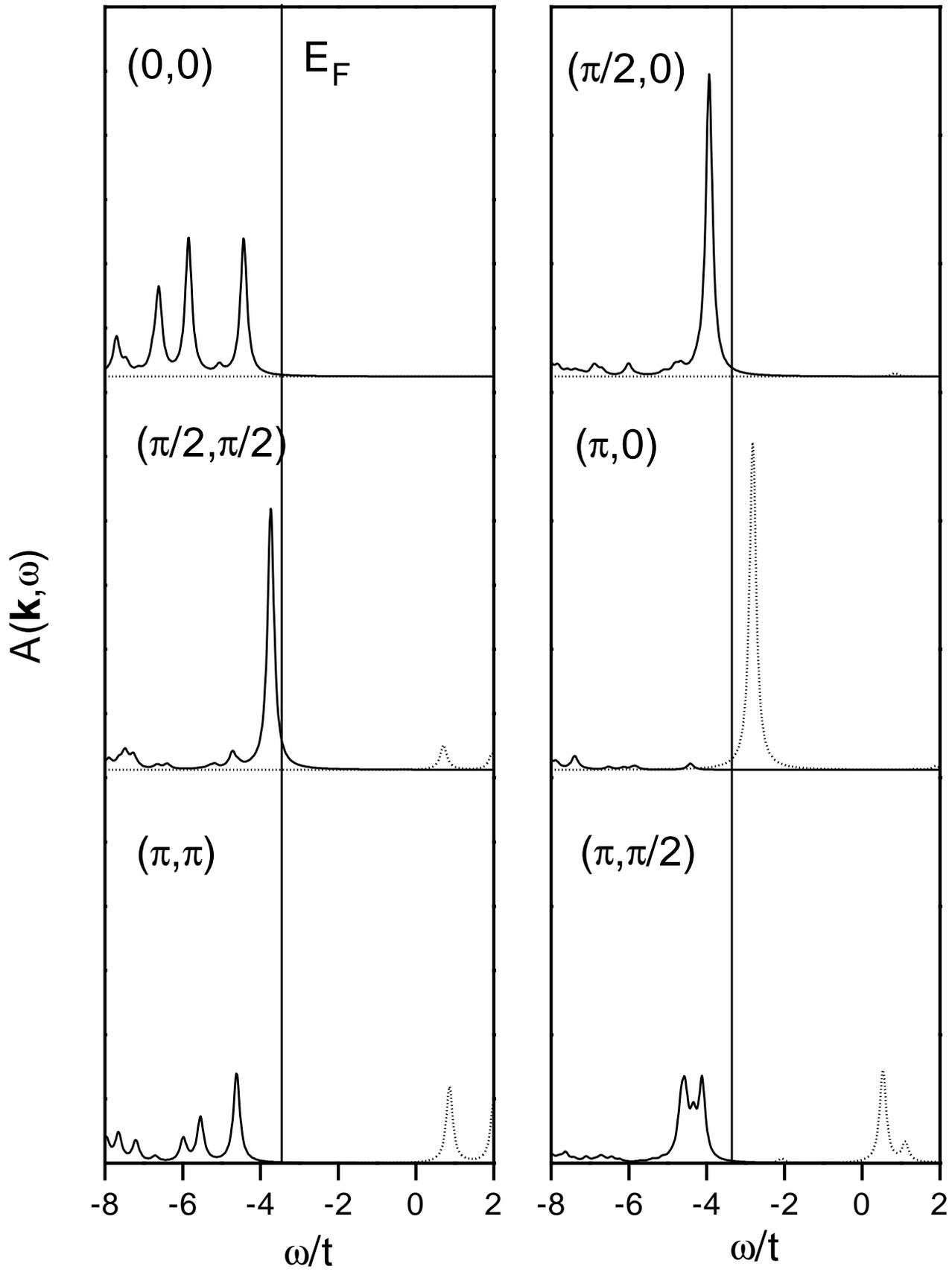
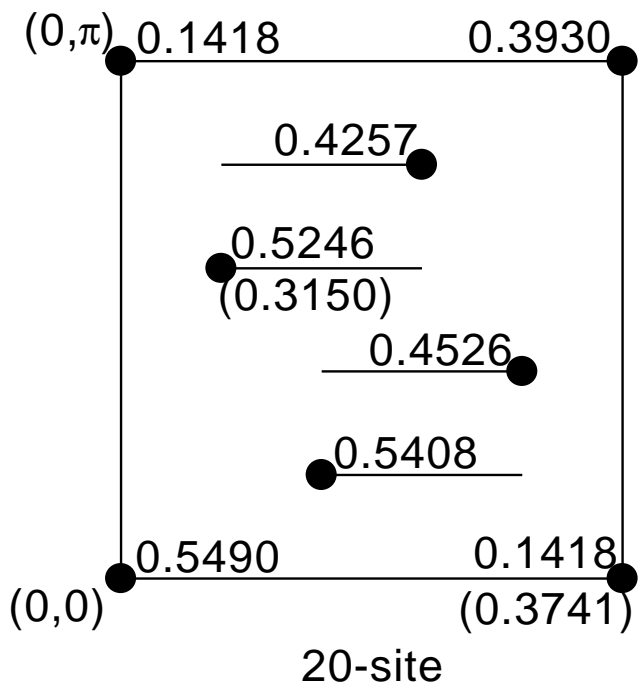
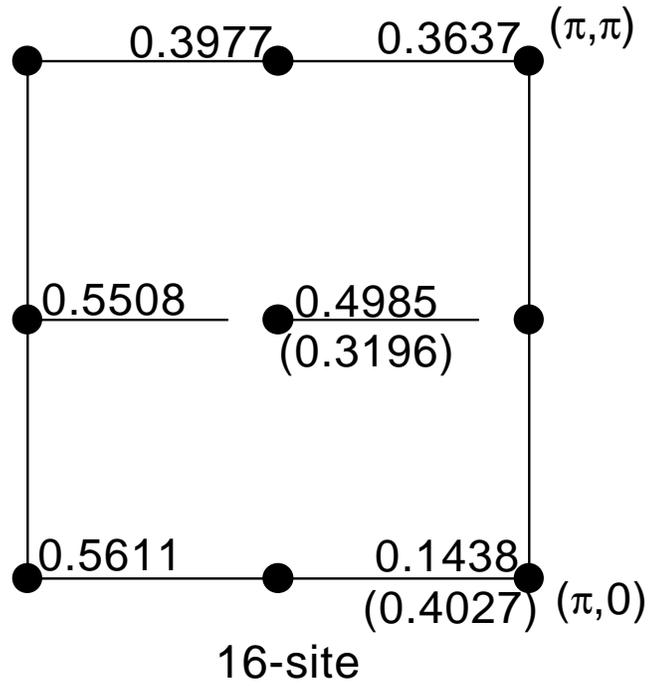
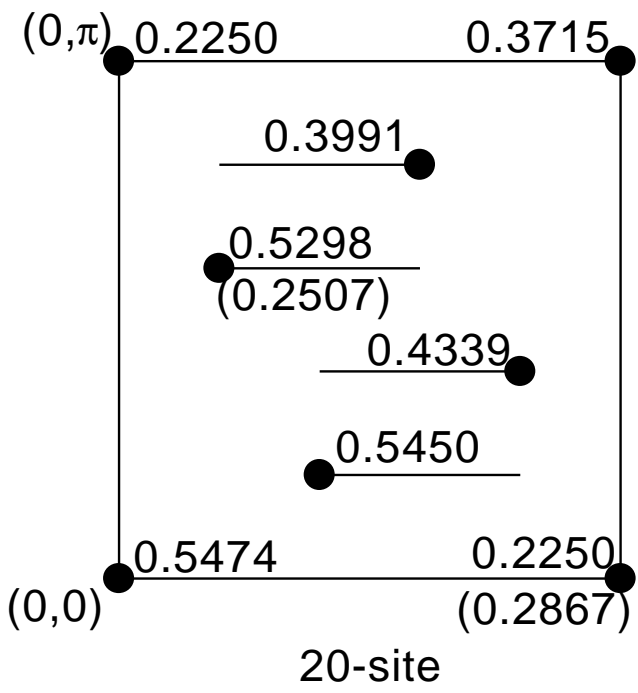
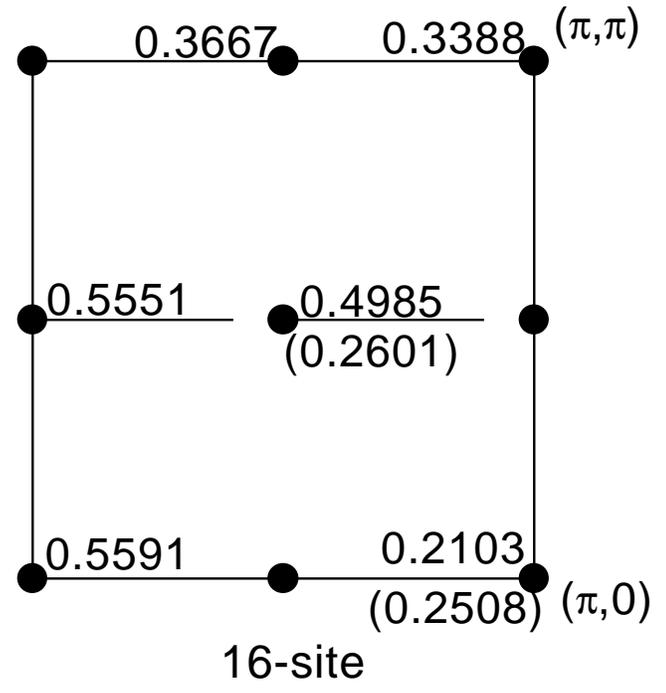
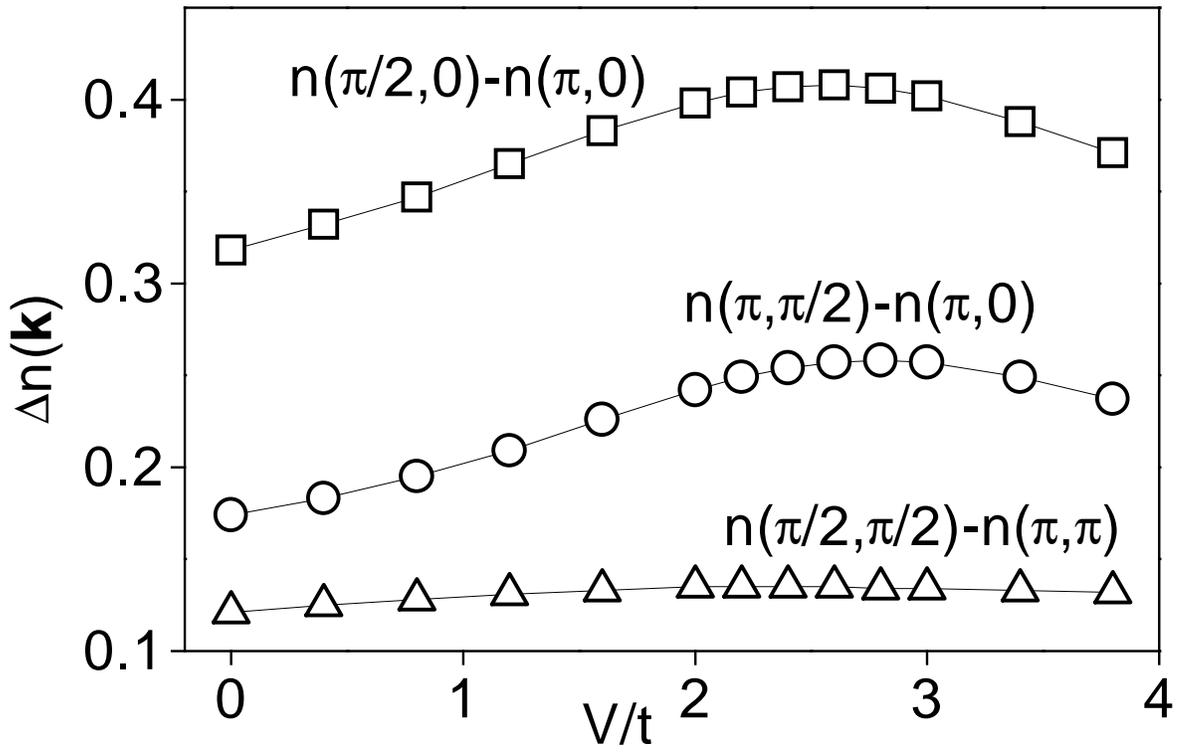
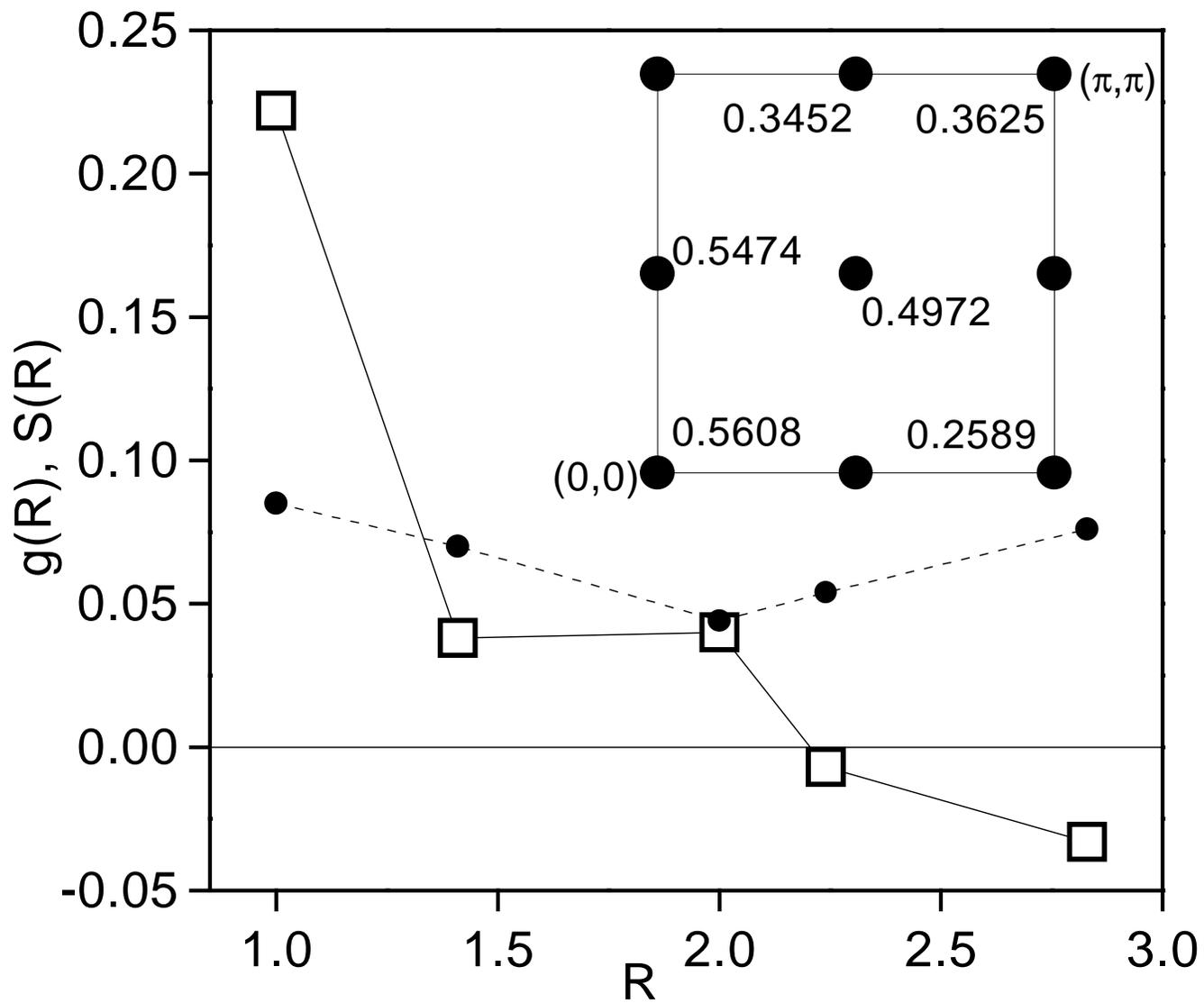
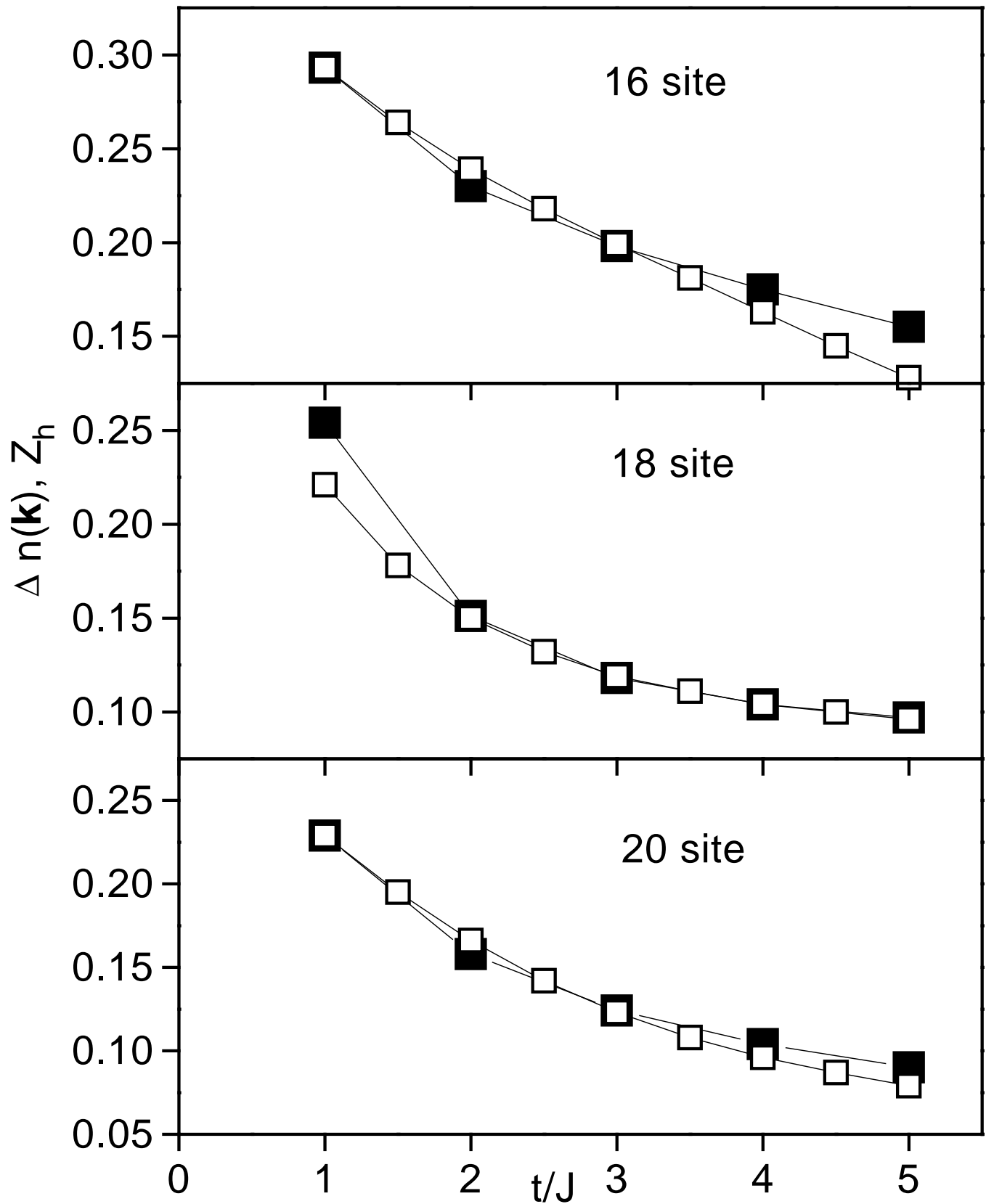
\end{document}